\documentclass[a4paper,10pt,twoside]{cpc-hepnp}

\usepackage{multicol}
\usepackage{graphicx}
\usepackage{booktabs}
\usepackage{amssymb,bm,mathrsfs,bbm,amscd}
\usepackage[tbtags]{amsmath}
\usepackage{lastpage}

\begin{document}

\fancyhead[c]{\small Chinese Physics C~~~Vol. xx, No. x (201x) xxxxxx}
\fancyfoot[C]{\small 010201-\thepage}

\footnotetext[0]{Received 14 March 2009}

\title{Analysis of the Energy Spectra of Ground States of Deformed Nuclei in rare-earth region
\thanks{Supported by IIUM and MOHE Fundamental Research Grant Scheme }}

\author{%
      A. A. Okhunov$^{1,3;1)}$\email{aokhunov@yahoo.com}%
\quad G. I. Turaeva$^{1}$ \quad M. U. Khandaker$^{2}$ \quad Noora
B. Rosli$^{2}$   } \maketitle

%%%\author{%
%%%      A. A. Okhunov$^{1,3;1)}$\email{aokhunov@yahoo.com}%
%%%\quad G. I. Turaeva$^{1,2;2)}$\email{hepnp@mail.ihep.ac.cn}%
%%%\quad M. U. Khandaker$^{1}$ \quad F. Jone$^{2}$ } \maketitle

\address{%
$^1$ Department of Science in Engineering, International Islamic
University Malaya, Kuala Lumpur 50728,
Malaysia\\
$^2$ Department of Physics, University of Malaya,
Kuala Lumpur 50603, Malaysia\\
$^3$ Institute of Nuclear Physics, Academy of Science Republic of
Uzbekistan, Tashkent 100214, Uzbekistan\\
}

\begin{abstract}
The $_{62}Sm$, $_{64}Gd$, $_{64}Dy$, $_{70}Yb$, $_{72}Hf$ and
$_{74}W$ nuclei are classified as deformed nuclei. Low-lying bands
are one of the most fundamental excitation modes in the energy
spectra of deformed nuclei. In this paper a theoretical analysis
of the experimental data within the phenomenological model is
presented. The energy spectra of ground states are calculated. It
is found the low-lying spectra of ground band states are in good
agreement with the experimental data.
\end{abstract}


%%%\begin{pacs}
%%%1---3 PACS(Physics and Astronomy Classification Scheme,
%%%http://www.aip.org/pacs/pacs.html/)
%%%\end{pacs}

\footnotetext[0]{\hspace*{-3mm}\raisebox{0.3ex}{$\scriptstyle\copyright$}2013
Chinese Physical Society and the Institute of High Energy Physics
of the Chinese Academy of Sciences and the Institute
of Modern Physics of the Chinese Academy of Sciences and IOP Publishing Ltd}%

\begin{multicols}{2}

\section{Introduction}

Though the structure of rotational deformed nuclei and the nature
of low excited levels in these nuclei studying already is 50
years, but these problem is rather actual and for today.

Nowadays, accumulated of rich experimental data on the excited and
low excited states of deformed nuclei, which is need to
development of their theoretical researches.

Phenomenological nuclear adiabatic model described by Bohr and
Mottelson \cite{Bohr} has been played a big role in understanding
the properties of deformed nuclei. According to this model, low
excitation states of even-even deformed nuclei are connected with
rotation of axial-symmetric nucleus as a whole. Such a simple
phenomenological explanation allows a description of a large
number of experimental data on deformed nuclei and predicts many
new properties of these nuclei.

In the present paper the deviations from the adiabatic theory
which appear in energies of ground band states is analyzed, within
the phenomenological model \cite{Usmanov97,Usmanov10}, which into
account Coriolis mixture of low-lying state bands. The objects of
calculation are $^{152-156}Sm$, $^{156-166}Gd$, $^{156-166}Dy$,
$^{166-176}Yb$, $^{170-180}Hf$ and $^{174-184}W$, isotopes. These
nuclei have been quite well studied experimentally such as in
nuclear reactions and Coulomb excitation
\cite{Reich03,Helmer96,Reich05}. Experiments make the systematical
measurement of properties of low-lying states.

\section{Energy spectra of $gr$- state bands}

The energy of rotational core $E_{rot}(I)$ is in agreement with
the energy of the ground state of rotational bands of even-even
deformed nuclei in the lower value of spin $I$.

The effective angular frequency of rotating nucleus is defined as
follows:

%\begin{footnotesize}
\begin{eqnarray}
\label{eq1}
%\begin{equation}
\omega_{eff}=\frac{E^{exp.}(I+1)-E^{exp.}(I-1)}{2}
%\end{equation}
\end{eqnarray}
%\end{footnotesize}

and hence the effective moment of inertia for states
$\Im_{eff}(I)$ in terms of the angular frequency of rotation
$\omega_{rot}(I)$ is

%\begin{footnotesize}
\begin{eqnarray}
\label{eq2}
%\begin{equation}
\Im_{eff}=\frac{\left(I(I+1)\right)^{1/2}}{\omega_{eff}(I)}.
%\label{Eq.2}
%\end{equation}
\end{eqnarray}
%\end{footnotesize}

From ~``Eq.~(\ref{eq2})'', we can calculate the effective moment
of inertia $\Im_{eff}(I)$. Nuclear angular frequency of rotation
$\omega_{eff}(I)$ is obtained by ~``Eq.~(\ref{eq1})'', with the
energy $E^{exp.}(I)$ from experiment \cite{Begzhanov89}. At low
angular frequency of rotation, i.e. in low spin $I\leq 8\hbar$ the
dependency is linear. We parameterize this dependency as follows:

%\begin{footnotesize}
\begin{eqnarray}
\label{eq3}
%\begin{equation}
\Im_{eff}=\Im_0+\Im_1\omega_{eff}^2(I).
%\label{Eq.3}
%\end{equation}
\end{eqnarray}
%\end{footnotesize}

\noindent Equation ~``Eq.~(\ref{eq3})'' defines the parameters of
inertia $\Im_0$ and $\Im_1$, for the effective moment of inertia
$\Im_{eff}(I)$ when $I\leq 8\hbar$.

The energy of the rotational ground band state is calculated using
the parameters $\Im_0$ and $\Im_1$ by Harris parametrization for
the energy and angular momentum \cite{Harris65}.

%\begin{footnotesize}
\begin{eqnarray}
\label{eq4}
%\begin{equation}
E_{rot}(I)=\frac{1}{2}\Im_0\omega_{rot}^2(I)+\frac{3}{4}\Im_1\omega_{rot}^4(I).
%\label{Eq.4}
%\end{equation}
\end{eqnarray}
%\end{footnotesize}

\section{Result and Discussion}

To analyze the properties of states of the ground band in deformed
nuclei, the phenomenological model of \cite{Usmanov97} has been
utilized.

The Numerical values for the parameters $\Im_0$ and $\Im_1$ are
determined using the least square method in ~``Eq.~(\ref{eq3})''.
These results are shown in Table~\ref{tab1}, for the isotopes
$^{152,154}Sm$, $^{158-162}Dy$ and $^{172-176}Yb$ respectively
$(\Im_0$ in $MeV^-1$ and  in $MeV^-3)$.

\begin{center}
\tabcaption{ \label{tab1}  The values of parameters $\Im_0$,
$\Im_1$ and $E_{2^+}$ ($MeV$) for isotopes $Sm$, $Gdy$, $Dy$,
$Yb$, $Hf$ and $W$, respectively.} \footnotesize
\begin{tabular*}{85mm}{c|c|c|c|c|r}
%%%%\begin{tabular*}{80mm}{c@{\extracolsep{\fill}}|c|c|c|c|c|r}
\toprule $A$ & $\Im_0$ & $\Im_1$ &  $E_{2^+}^{exp.}$ \ & $E_{2^+}^{theory}$  & $Q_0$ Ref.~\citep{Begzhanov89} \\
\hline
 \multicolumn{6}{c}{$Sm$}\\ \cline{1-6}
  \ 152 \ & 24.74 \ & 256.57 \ & 0.1218 \ &  0.1234 \ & 5.83 (057) \ \\
  \ 154 \ & 36.07 \ & 178.88 \ & 0.0724 \ &  0.0841 \ & 6.53 (15) \ \\
  \ 156 \ & 39.22 \ &  98.36 \ & 0.0760 \ &  0.0778 \ & 7.85 (79) \ \\  \hline
\multicolumn{6}{c}{$Gd$}\\ \cline{1-6}
 \ 156 \ & 38.74 \ &  95.25 \ & 0.0890 \ & 0.0769 \ & 6.76 (34) \ \\
 \ 158 \ & 37.52 \ & 107.00 \ & 0.0795 \ & 0.0795 \ & 7.03 (4) \ \\
 \ 160 \ & 39.72 \ &  83.49 \ & 0.0755 \ & 0.0752 \ & 7.16 (2) \ \\   \hline
\multicolumn{6}{c}{$Dy$}\\ \cline{1-6}
 \ 156 \ & 21.93 \ & 238.14 \ & 0.1370 \ & 0.1290 \ & 6.12 (6) \ \\
 \ 158 \ & 29.69 \ & 174.26 \ & 0.0989 \ & 0.0990 \ & 6.85 (8) \ \\
 \ 160 \ & 33.96 \ & 131.07 \ & 0.0898 \ & 0.0870 \ & 6.91 (20) \ \\
 \ 162 \ & 36.61 \ & 105.77 \ & 0.0806 \ & 0.3350 \ & 7.13  \ \\
 \ 164 \ & 40.25 \ & 121.09 \ & 0.0734 \ & 0.0740 \ & 7.49  \ \\
 \ 166 \ & 38.68 \ &  73.82 \ & 0.0766 \ & 0.0770 \ & --  \ \\ \hline
\multicolumn{6}{c}{$Yb$}\\ \cline{1-6}
 \ 166 \ & 28.75 \ & 132.20 \ & 0.1024 \ & 0.1032 \ & 7.26 (14) \ \\
 \ 168 \ & 33.34 \ & 148.77 \ & 0.0877 \ & 0.0890 \ & 7.62 (14) \ \\
 \ 170 \ & 35.06 \ &  87.82 \ & 0.0843 \ & 0.0850 \ & 7.80 (30) \ \\
 \ 172 \ & 37.60 \ &  81.31 \ & 0.0788 \ & 0.0790 \ & 7.91 (18) \ \\
 \ 174 \ & 38.71 \ &  76.65 \ & 0.0765 \ & 0.0770 \ & 7.82 (24) \  \\
 \ 176 \ & 36.00 \ &  63.96 \ & 0.0821 \ & 0.0830 \ & 7.59 (3)  \ \\  \hline
\multicolumn{6}{c}{$Hf$}\\ \cline{1-6}
 \ 170 \ & 27.93 \ & 248.95 \ & 0.1008 \ & 0.1007 \ & 7.14 (30) \ \\
 \ 172 \ & 30.02 \ & 172.45 \ & 0.0953 \ & 0.0953 \ & 6.87 (18) \ \\
 \ 174 \ & 31.09 \ & 134.03 \ & 0.0900 \ & 0.0901 \ & 7.29 (24) \ \\
 \ 176 \ & 33.70 \ &  92.27 \ & 0.0884 \ & 0.0877 \ & 7.23 (8) \ \\
 \ 178 \ & 31.95 \ &  71.88 \ & 0.0932 \ & 0.0928 \ & 6.98 (4) \  \\
 \ 180 \ & 32.06 \ &  36.53 \ & 0.0933 \ & 0.0931 \ & 6.93 (3)  \ \\   \hline
\multicolumn{6}{c}{$W$}\\ \cline{1-6}
 \ 174 \ & 25.06 \ & 212.67 \ & 0.1130 \ & 0.1136 \ & -- \ \\
 \ 176 \ & 25.91 \ & 177.61 \ & 0.1091 \ & 0.1097 \ & -- \ \\
 \ 178 \ & 26.87 \ & 143.47 \ & 0.1061 \ & 0.1069 \ & --  \ \\
 \ 180 \ & 27.51 \ & 127.14 \ & 0.1036 \ & 0.1030 \ & 6.24 (11) \ \\
 \ 182 \ & 28.31 \ & 110.50 \ & 0.1001 \ & 0.1045 \ & 6.57 (8) \  \\
 \ 184 \ & 29.22 \ &  98.36 \ & 0.1112 \ & 0.1015 \ & 6.27 (8)  \ \\
      \hline
\end{tabular*}
\vspace{0mm}
\end{center}
\vspace{0mm}
\end{multicols}

The values of angular frequency and energy spectra of $Sm$, $Dy$
and $Yb$ is illustrated in Table~\ref{tab2}, respectively. All of
the predicted data seems with agreed with the experimental data.

\begin{center}
\tabcaption{ \label{tab2}  The Value of angular frequency and
energy spectra of the $Sm$ ($E \ in \ MeV$).} \footnotesize
%%%%%\tabcaption{\label{tab2}  Wide table.} \footnotesize
\begin{tabular*}{175mm}{@{\extracolsep{\fill}}c|c|c|c|c|c|c|c|c|c}
\hline
 \ $I$ & \multicolumn{3}{c|}{$^{152}Sm$}& \multicolumn{3}{c|}{$^{154}Sm$} & \multicolumn{3}{c}{$^{156}Sm$}\\ \cline{2-10}
  & \ $\omega_{rot}^{theor}(I)$  \ & \ $E_{rot}^{exp.}(I)$ \ &
  \ $E_{rot}^{theor}(I)$ \ & \ $\omega_{rot}^{theor}(I)$  \ & \ $E_{rot}^{exp.}(I)$ \ &
  \ $E_{rot}^{theor}(I)$ \ & \ $\omega_{rot}^{theor}(I)$  \ & \ $E_{rot}^{exp.}(I)$ \ &
  \ $E_{rot}^{theor}(I)$ \  \\ \hline
  2$^+$ & 0.091 \ & 0.122 \ & 0.116 \ & 0.066 \ & 0.082 \ &
  \ 0.082 & 0.062 \ & 0.076 \ & 0.076   \\
  4$^+$ & 0.147 \ & 0.366 \ & 0.360 \ & 0.116 \ & 0.267 \ &
  \ 0.268 & 0.111 \ & 0.250 \ & 0.251   \\
  6$^+$ & 0.190 \ & 0.707 \ & 0.701 \ & 0.160 \ & 0.544 \ &
  \ 0.546 & 0.156 \ & 0.517 \ & 0.519   \\
  8$^+$ & 0.225 \ & 1.126 \ & 1.119 \ & 0.197 \ & 0.903 \ &
  \ 0.904 & 0.197 \ & 0.872 \ & 0.874   \\
 10$^+$ & 0.254 \ & 1.609 \ & 1.599 \ & 0.230 \ & 1.333 \ &
  \ 1.333 & 0.235 \ & 1.307 \ & 1.307   \\
 12$^+$ & 0.280 \ & 2.149 \ & 2.134 \ & 0.260 \ & 1.826 \ &
  \ 1.824 & 0.269 \ & 1.819 \ & 1.812   \\
   \hline\hline
 \ $I$ & \multicolumn{3}{c|}{$^{156}Dy$}& \multicolumn{3}{c|}{$^{158}Dy$} & \multicolumn{3}{c}{$^{160}Dy$}\\ \cline{1-10}
  2$^+$ & 0.101 \ & 0.138 \ & 0.129 \ & 0.080 \ & 0.099 \ &
  \ 0.100 & 0.071 \ & 0.087 \ & 0.087   \\
  4$^+$ & 0.160 \ & 0.404 \ & 0.396 \ & 0.136 \ & 0.317 \ &
  \ 0.319 & 0.124 \ & 0.284 \ & 0.286   \\
  6$^+$ & 0.204 \ & 0.770 \ & 0.763 \ & 0.183 \ & 0.638 \ &
  \ 0.640 & 0.171 \ & 0.581 \ & 0.584   \\
  8$^+$ & 0.239 \ & 1.216 \ & 1.207 \ & 0.222 \ & 1.044 \ &
  \ 1.046 & 0.213 \ & 0.967 \ & 0.970   \\
 10$^+$ & 0.268 \ & 1.725 \ & 1.716 \ & 0.255 \ & 1.520 \ &
  \ 1.525 & 0.249 \ & 1.429 \ & 1.433   \\
 12$^+$ & 0.294 \ & 2.286 \ & 2.280 \ & 0.285 \ & 2.049 \ &
  \ 2.067 & 0.282 \ & 1.951 \ & 1.965   \\
  \hline
\ & \multicolumn{3}{c|}{$^{162}Dy$}&
\multicolumn{3}{c|}{$^{164}Dy$} & \multicolumn{3}{c}{$^{166}Dy$}\\
\cline{1-10}
  2$^+$ & 0.066 \ & 0.081 \ & 0.081 \ & 0.060 \ & 0.073 \ &
  \ 0.074 & 0.063 \ & 0.077 \ & 0.077   \\
  4$^+$ & 0.117 \ & 0.266 \ & 0.268 \ & 0.107 \ & 0.242 \ &
  \ 0.244 & 0.113 \ & 0.254 \ & 0.255   \\
  6$^+$ & 0.164 \ & 0.549 \ & 0.551 \ & 0.151 \ & 0.501 \ &
  \ 0.504 & 0.160 \ & 0.527 \ & 0.530   \\
  8$^+$ & 0.206 \ & 0.921 \ & 0.924 \ & 0.190 \ & 0.844 \ &
  \ 0.846 & 0.203 \ & 0.892 \ & 0.894   \\
 10$^+$ & 0.244 \ & 1.375 \ & 1.376 \ & 0.226 \ & 1.261 \ &
  \ 1.264 & 0.244 \ & 1.341 \ & 1.342   \\
 12$^+$ & 0.279 \ & 1.901 \ & 1.900 \ & 0.258 \ & 1.745 \ &
  \ 1.749 & 0.281 \ & 1.868 \ & 1.867   \\
    \hline\hline
 \ $I$ & \multicolumn{3}{c|}{$^{166}Yb$}& \multicolumn{3}{c|}{$^{168}Yb$} & \multicolumn{3}{c}{$^{170}Yb$}\\ \cline{1-10}
  2$^+$ & 0.083 \ & 0.102 \ & 0.103 \ & 0.072 \ & 0.088 \ &
  \ 0.089 & 0.069 \ & 0.084 \ & 0.085   \\
  4$^+$ & 0.142 \ & 0.331 \ & 0.332 \ & 0.125 \ & 0.287 \ &
  \ 0.289 & 0.123 \ & 0.277 \ & 0.280   \\
  6$^+$ & 0.193 \ & 0.668 \ & 0.670 \ & 0.172 \ & 0.585 \ &
  \ 0.589 & 0.172 \ & 0.574 \ & 0.577   \\
  8$^+$ & 0.235 \ & 1.098 \ & 1.100 \ & 0.212 \ & 0.970 \ &
  \ 0.975 & 0.217 \ & 0.964 \ & 0.967   \\
 10$^+$ & 0.272 \ & 1.606 \ & 1.610 \ & 0.247 \ & 1.426 \ &
  \ 1.435 & 0.257 \ & 1.438 \ & 1.442   \\
 12$^+$ & 0.305 \ & 2.176 \ & 2.188 \ & 0.278 \ & 1.936 \ &
  \ 1.962 & 0.293 \ & 1.984 \ & 1.993   \\
   \hline
\ & \multicolumn{3}{c|}{$^{172}Yb$}&
\multicolumn{3}{c|}{$^{174}Yb$} & \multicolumn{3}{c}{$^{176}Yb$}\\
\cline{1-10}
  2$^+$ & 0.065 \ & 0.079 \ & 0.079 \ & 0.063 \ & 0.076 \ &
  \ 0.077 & 0.067 \ & 0.082 \ & 0.083   \\
  4$^+$ & 0.116 \ & 0.260 \ & 0.262 \ & 0.113 \ & 0.253 \ &
  \ 0.255 & 0.121 \ & 0.272 \ & 0.274   \\
  6$^+$ & 0.163 \ & 0.540 \ & 0.543 \ & 0.159 \ & 0.526 \ &
  \ 0.529 & 0.171 \ & 0.565 \ & 0.568   \\
  8$^+$ & 0.207 \ & 0.912 \ & 0.914 \ & 0.203 \ & 0.890 \ &
  \ 0.892 & 0.217 \ & 0.955 \ & 0.958   \\
 10$^+$ & 0.247 \ & 1.370 \ & 1.368 \ & 0.243 \ & 1.336 \ &
  \ 1.339 & 0.260 \ & 1.431 \ & 1.437   \\
 12$^+$ & 0.283 \ & 1.907 \ & 1.899 \ & 0.279 \ & 1.861 \ &
  \ 1.862 & 0.299 \ & 1.985 \ & 1.998   \\
    \hline
\end{tabular*}%
\end{center}

The comparison between the calculated values of energy of the
ground band state with experimental data \cite{Begzhanov89} is
given for the isotopes $_{62}Sm$, $_{64}Gd$, $_{66}Dy$, $_{70}Yb$,
$_{72}Hf$ and $_{74}W$ in Figs.~\ref{fig1-fig5} Figs. 1--5,
respectively. From the figures, we see that energy difference
$\Delta E(I)=E^{theor}(I)-E^{exp.}(I)$, increases with the
increase in the angular momentum $I$. This is due to the
occurrence of the non-adiabaticity of energy rotational bands in
large spin.

\begin{center}
\includegraphics[width=12cm]{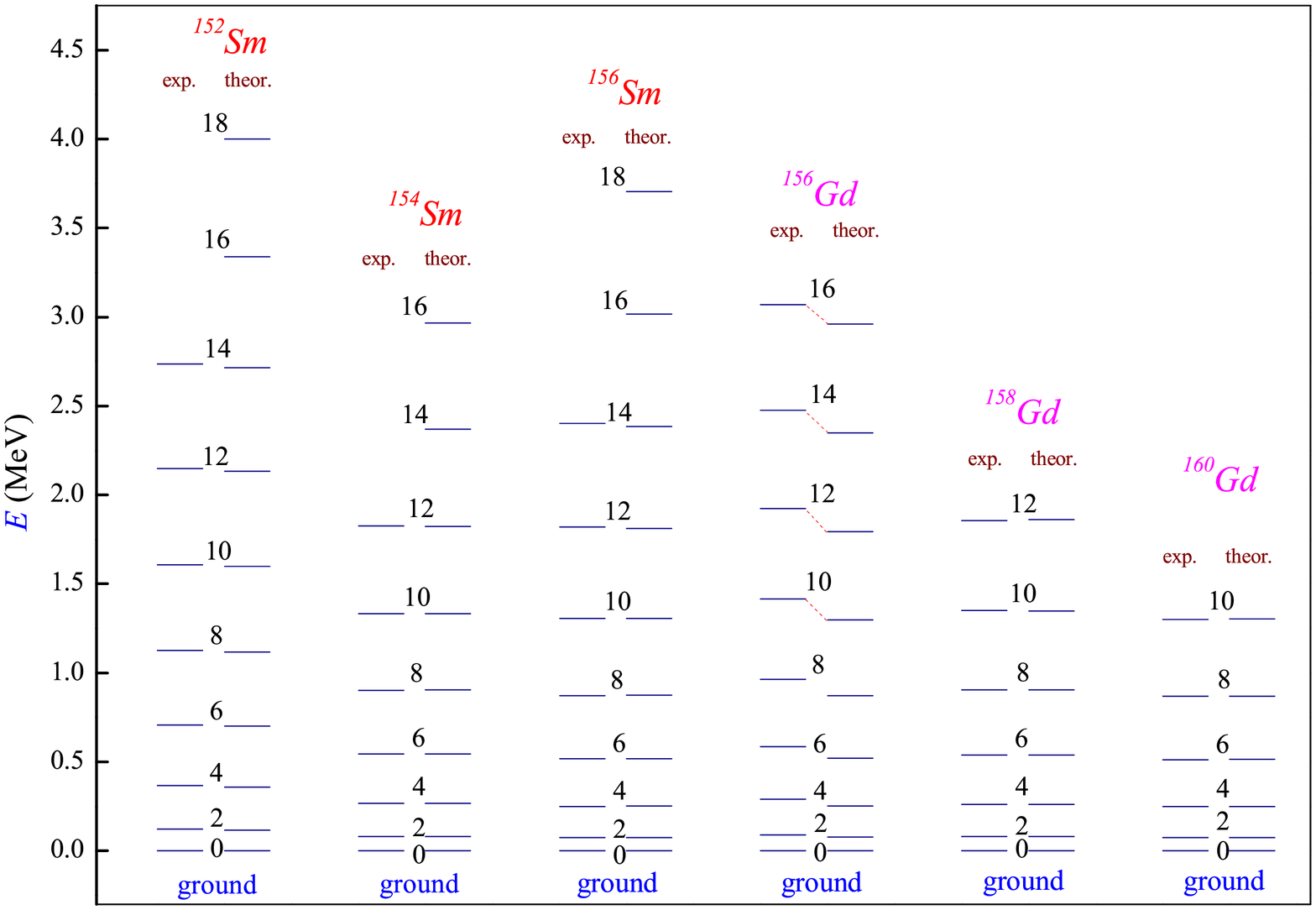}
\figcaption{\label{fig1} The comparison experimental and
theoretical spectra energy of ground bands for isotopes $Sm$ and
$Gd$, correspondingly.}
\end{center}

\begin{center}
\includegraphics[width=12cm]{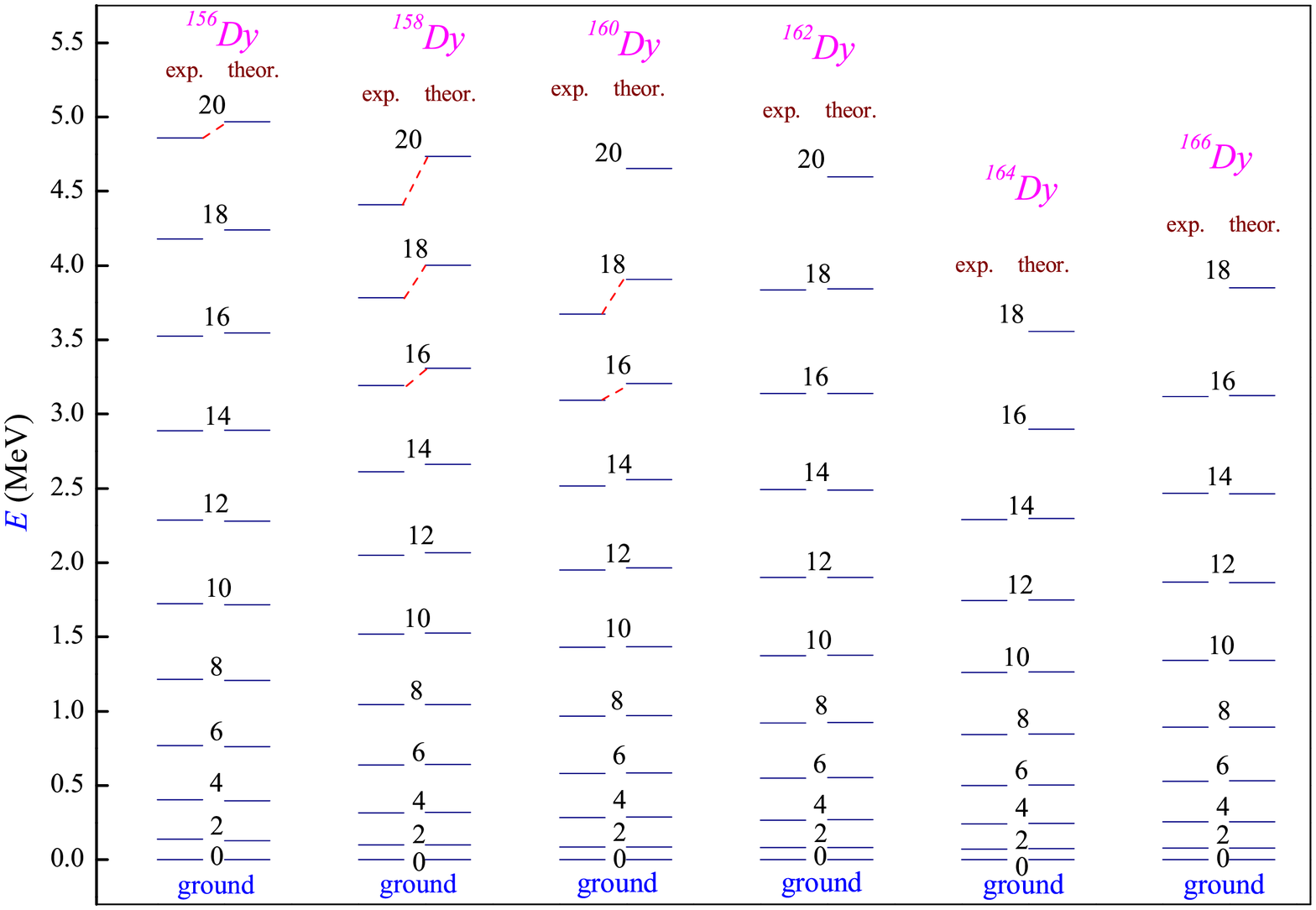}
\figcaption{\label{fig2} The comparison experimental and
theoretical spectra energy of ground bands for isotopes $Sm$ and
$Dy$, correspondingly.}
\end{center}

\begin{center}
\includegraphics[width=12cm]{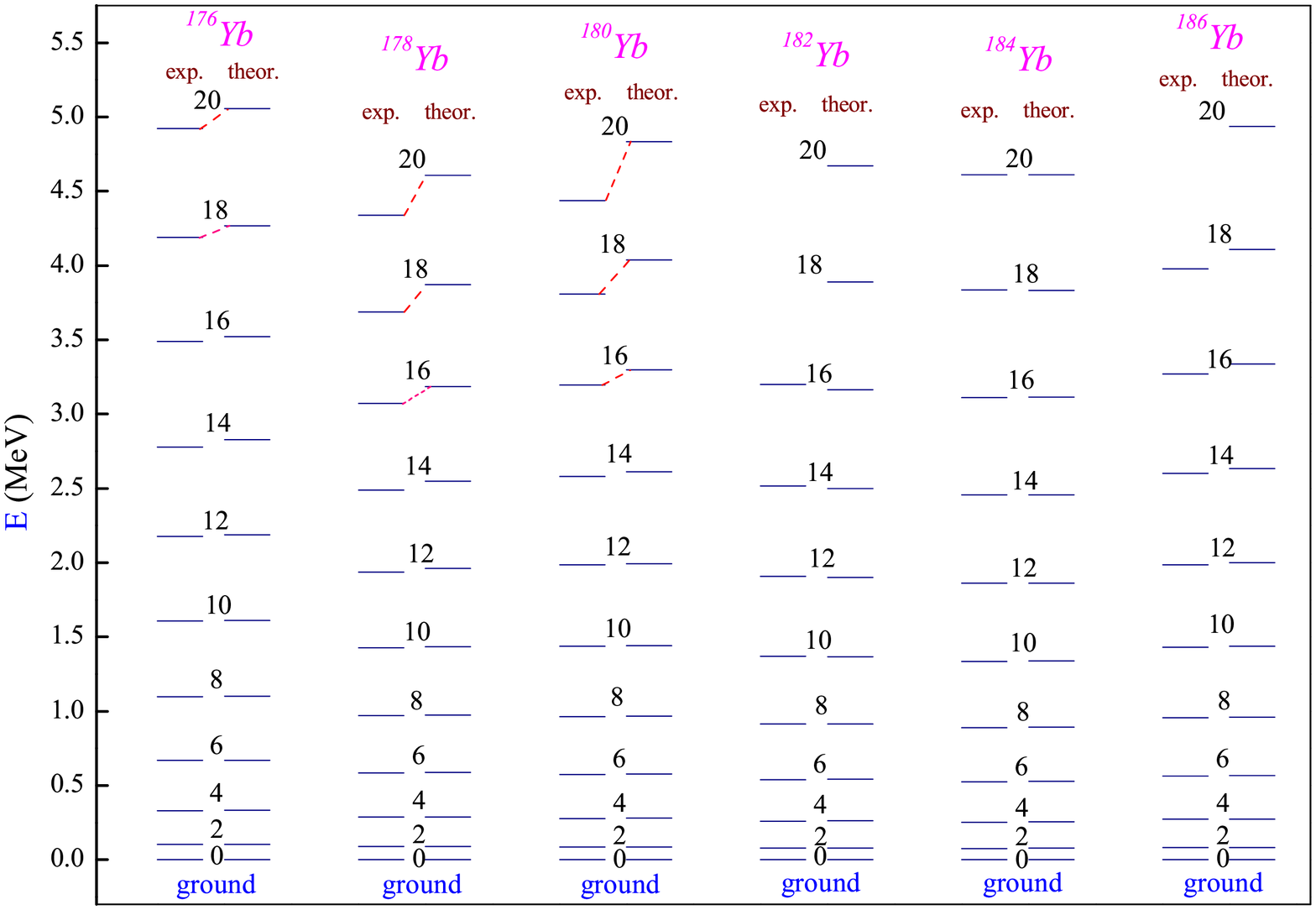}
\figcaption{\label{fig3} The comparison experimental and
theoretical spectra energy of ground bands for isotopes $Sm$ and
$Yb$, correspondingly.}
\end{center}

\begin{center}
\includegraphics[width=12cm]{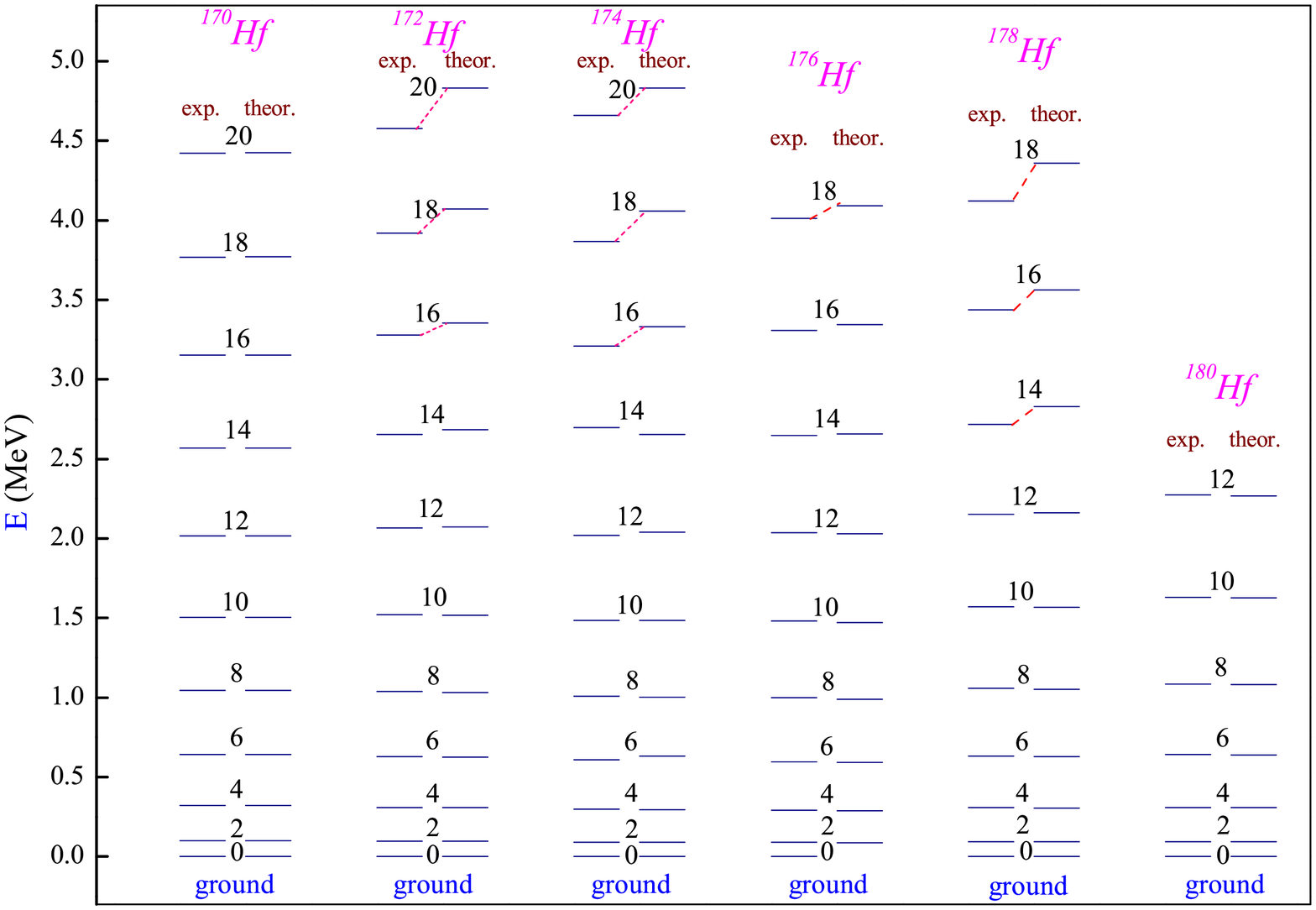}
\figcaption{\label{fig4} The comparison experimental and
theoretical spectra energy of ground bands for isotopes $Sm$ and
$Hf$, correspondingly.}
\end{center}

\begin{center}
\includegraphics[width=12cm]{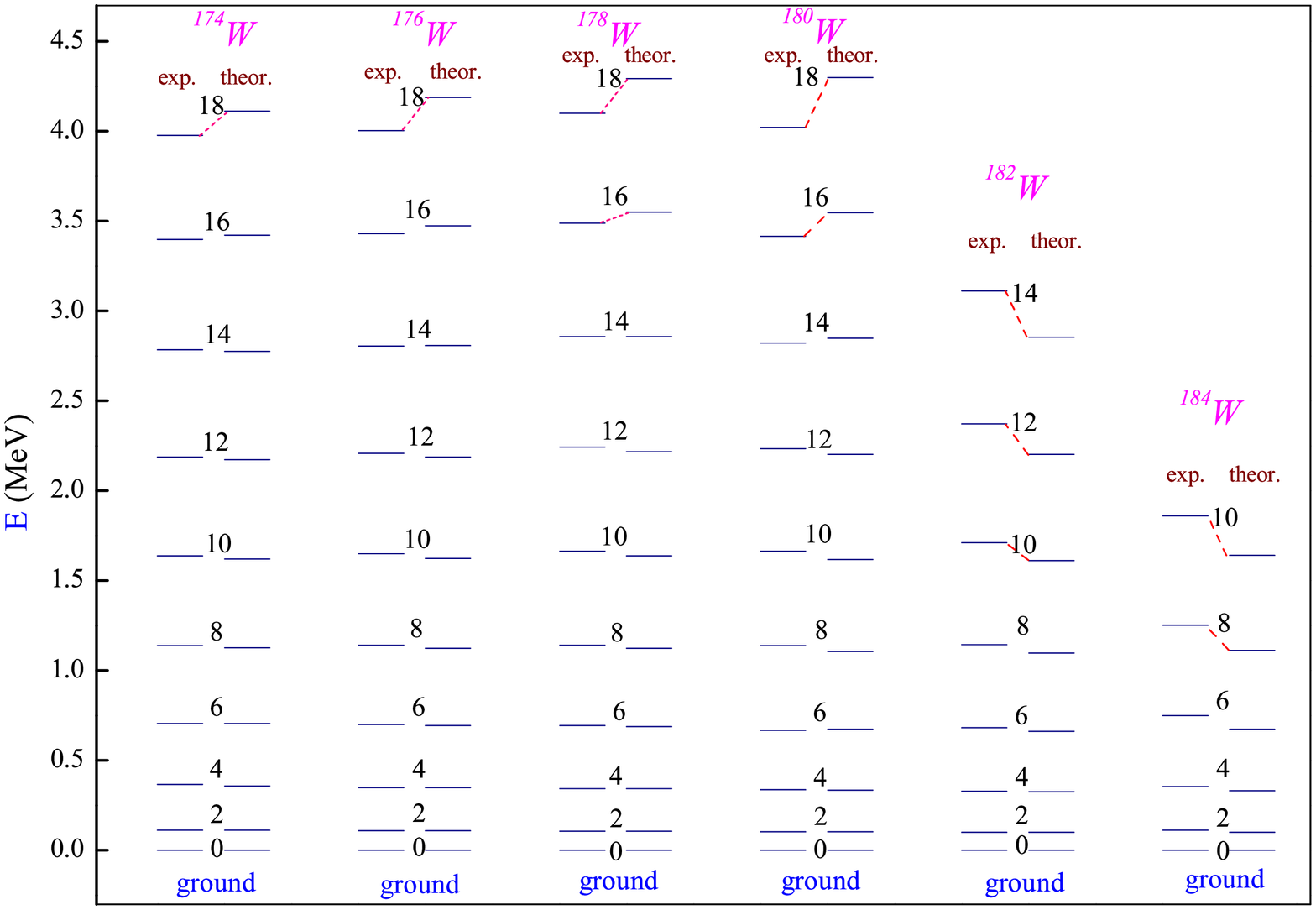}
\figcaption{\label{fig5} The comparison experimental and
theoretical spectra energy of ground bands for isotopes $Sm$ and
$W$, correspondingly.}
\end{center}

\begin{multicols}{2}

\section{ CONCLUSIONS AND FURTHER STUDIES}

This work is based on the phenomenological model \cite{Usmanov97,
Usmanov10}, which is clearly describes large number of
experimental data by the deviation properties of the positive
parity in even-even deformed nuclei from the role of adiabatic
theory. Spectral energy of ground states for the isotopes
$^{152-156}Sm$, $^{156-166}Gd$, $^{156-166}Dy$, $^{166-176}Yb$,
$^{172-182}Hf$ and $^{172-176}W$ were calculated show the
violation in the $E(I)\sim I(I+1)$ law. This is explained by the
fact that the nuclear core under rotation with the large mixture
frequency of ground-state bands with other rotational bands that
have vibrational characters. The calculation takes into account
the Coriolis mixing of positive parity states which has good
agreement with experimental data.

\section{ACKNOWLEDGMENTS}

\acknowledgments{We thank the IIUM University Research Grant Type
B (EDW B13-034-0919) and MOHE Fundamental Research Grant Scheme
(FRGS13-077-0315). The author A. A. Okhunov is grateful to Prof.
Ph.N. Usmanov for useful discussion and exchange ideas.}

\end{multicols}

\vspace{15mm}

\begin{multicols}{2}

\end{multicols}

\clearpage

\end{document}